\documentclass[12pt]{article}

\usepackage{graphicx}
\usepackage{setspace}

\begin{document}

\title{An omnidirectional retroreflector based on the transmutation of dielectric singularities}
\author{
Yun Gui Ma$^1$, C.\  K.\ Ong$^2$, Tom\'a\v{s} Tyc$^{3}$ and Ulf Leonhardt$^{4\ast}$\\
$^1$Temasek Laboratories, National University of Singapore,\\ Singapore 119260, Singapore\\
$^2$Centre for Superconducting and Magnetic Materials, \\
Department of Physics, National University of Singapore,\\ Singapore 117542, Singapore\\
$^3$Institute of Theoretical Physics and Astrophysics,\\
Masaryk University, Kotlarska 2, 61137 Brno, Czech Republic\\
$^4$School of Physics and Astronomy,\\ University of St Andrews, North Haugh, St Andrews, KY16 9SS,\\ UK}

\maketitle

\newpage

{\bf In the field of transformation optics$^{1-6}$, metamaterials$^{7-11}$ mimic the effect of coordinate transformations on electromagnetic waves, creating the illusion that the waves are propagating through a virtual space. Transforming space by appropriately designed materials makes devices possible that have been deemed impossible. In particular, transformation optics has led to the demonstration of invisibility cloaking for microwaves$^{12,13}$, surface plasmons$^{14}$ and infrared light$^{15,16}$. Here we report the achievement of another ``impossible task''. We implement, for microwaves, a device that would normally require a dielectric singularity, an infinity in the refractive index. We transmute$^{17}$ a singularity in virtual space into a mere topological defect in a real metamaterial. In particular, we demonstrate an omnidirectional retroreflector$^{18,19}$, a device for faithfully reflecting images and for creating high visibility, from all directions. Our method is robust, potentially broadband and similar techniques could be applied for visible light.}

Dielectric singularities are points where the refractive index $n$ reaches infinity or zero, where electromagnetic waves travel infinitely slow or infinitely fast. Such singularities cannot be made in practice for a broad spectral range, but one can transmute them into topological defects of anisotropic materials$^{17}$. Here is a brief summary of the underlying theoretical results$^{17}$: imagine an isotropic index profile with a singularity in virtual space. Suppose that around the singularity the index profile is spherically symmetric, described by the function $n (r')$ in spherical coordinates $\{r', \vartheta', \phi'\}$. We use primes to distinguish virtual space from real space. Then we represent the virtual coordinates in real space $\{r, \vartheta, \phi\}$ by $\vartheta'=\vartheta$, $\phi'=\phi$ and the continuous function $r' (r)$. We require that  $r' (r)$ obeys
\begin{equation}
n (r') \frac{{\rm d} r'}{{\rm d} r} = n_0
\quad \mbox{or, equivalently,}\quad
r = r(r') = \frac{1}{n_0} \int n(r')\,{\rm d} r'
\label{condition}
\end{equation}
where $n_0$ is a constant chosen such that beyond some radius $a$ the virtual and the real coordinates coincide, $r'=r$ for $r\ge a$. The radius $a$ defines the boundary of the device. Theory$^{4,17}$ shows that both the coordinate transformation and the virtual index profile are implemented by a material with the dielectric tensors
\begin{equation}
\varepsilon^i_{\ j} = \mu^i_{\ j} = {\rm diag} \Big( \frac{n^2 r'^2}{n_0 r^2}, n_0, n_0   \Big)
\label{epsmu}
\end{equation}
in spherical coordinates $x^i = \{r, \vartheta, \phi\}$. The tensors of the electric permittivity$^{20}$ $\varepsilon^i_{\ j}$ and the magnetic permeability$^{20}$ $\mu^i_{\ j}$ describe an impedance--matched$^{20}$, anisotropic dielectric$^{20}$ with varying radial component. Such media are said to be anisotropic, because they respond to different electromagnetic--field components differently, as described by the tensors (\ref{epsmu}), although the device they constitute is spherically symmetric. As real space and virtual space coincide for $r\ge a$, the device with the properties (\ref{epsmu}) has the same physical effect as the index profile $n$, but the dielectric tensors remain finite, as long as $n (r')$ does not diverge faster than $r'^{-1}$. The direction of dielectric anisotropy, though, is not defined at $r=0$, the position of the virtual singularity: the singularity has been transformed into a topological defect. 

Consider, for example, the Eaton lens$^{18,19}$ illustrated in Fig.\ 1. The Eaton lens would reflect light back to where it came from, while faithfully preserving any image the light carries (apart from inverting the image). The lens is spherically symmetric, and so it would retroreflect light regardless of direction: it would make a perfect omnidirectional retroreflector. In contrast, conventional optical retroreflectors --- ``cat's eyes'' --- are made of mirrors and have a finite acceptance angle. A metallic sphere is a omnidirectional retroreflector as well, but only for rays that directly hit its centre. The sphere scatters off-centre light and distorts images, and so does a metallic rod in planar illumination. 
The Eaton lens is characterized by the index profile$^{18}$
\begin{equation}
n = \sqrt{\frac{2 a}{r'} -1} \quad {\rm for} \,\, r'<a \quad
{\rm and} \quad n=1 \quad {\rm for} \,\, r'\geq a   \,.
\label{eaton}
\end{equation}
To understand why the profile (\ref{eaton}) acts as a perfect retroreflector one can use the following analogy$^{21}$: a light ray corresponds to the trajectory of a fictitious Newtonian particle that moves with energy $E$ in the potential $U$ where $U-E=-n^2/2$ (in dimensionless units). The Eaton lens corresponds to the Kepler potential$^{21}$ $U=-a/r'$ for $r'<a$ and $U=-1$ outside of the device, with total energy $E=-1/2$. The fictitious particle draws a half Kepler ellipse around the centre of attraction. So it leaves in precisely the opposite direction it came from$^{19}$: light is retroreflected.
 
One sees that the profile (\ref{eaton}) diverges with the power $-1/2$ and therefore an Eaton lens has never been made. However, we can transmute the singular isotropic profile into the regular but anisotropic material with the properties (\ref{epsmu}) by the coordinate transformation
\begin{equation}
r = \frac{2 a}{n_0} \left[ \arcsin \sqrt{\frac{r'}{2 a}} + \sqrt{ \frac{r'}{2 a} \left( 1 - \frac{r'}{2 a} \right)} \right]  \,,\quad n_0 = 1 + \frac{\pi}{2}
\end{equation}
for $r'<a$ and $r=r'$ for $r'>a$, because this $r (r')$ solves the integral (\ref{condition}) and approaches $r'$ at $r'=a$. The singularity of the Eaton lens has been transformed away.

We found that this transmutation of singularities is remarkably robust: Figures 2a and 2b compare numerical simulations of electromagnetic wave propagation in an ideal and a transmuted cylindrical Eaton lens. The cylindrical lens of Fig.\ 1b is characterized by its diagonal dielectric tensors in cylindrical coordinates $x^i=\{r, \varphi, z\}$, but we use the same functions (\ref{epsmu}) for the diagonal components  $\varepsilon_i$ and $\mu_i$ as in the case of the spherical lens,
\begin{equation}
\varepsilon_z = \mu_{\varphi} = n_0  \,,\quad \mu_r = \frac{n^2 r'^2}{n_0 r^2} \,.
\label{cylinder}
\end{equation}
We study the propagation of polarized electromagnetic waves in the $\{r,\varphi \}$ plane with the electric field pointing in $z$ direction. Such waves are not influenced by the other non--zero tensor components $\varepsilon_r$, $\varepsilon_\varphi$ and $\mu_z$. As Figs.\ 2a and 2b show, the wave propagations in the two devices are nearly identical, although the theory (\ref{epsmu}) presumes a spherical transformation, not a cylindrical one. Moreover, the corresponding biscattering diagrams indicate that the transformed medium appears as a better retroreflector than the numerically simulated Eaton lens, because of numerical problems associated with the singularity in the Eaton profile (\ref{eaton}). Furthermore, Fig.\ 2c shows the effect of another modification of the ideal case (\ref{epsmu}) that is useful in practice: in this simulation we rescale the dielectric functions in the device, but not in the air around it. Maxwell's equations are not changed if one rescales the dielectric functions (\ref{cylinder}) by the constant $n_0$ as
\begin{equation}
\varepsilon_z = n_0^2  \,,\quad \mu_{\varphi} = 1 \,,\quad \mu_r = \frac{n^2 r'^2}{n_0^2 r^2}   \,.
\label{rescaled}
\end{equation}
Consequently, the wave propagation in such media is the same as in materials with the dielectric functions (\ref{cylinder}). However, in reality the rescaling can only be done in the device, but not in air where $\varepsilon=\mu=1$. So the impedance$^{20}$ is mismatched at the interface, which causes reflections$^{20}$ --- scattering --- but, as Fig.\ 2c shows, the phase profile of the outgoing wave is preserved, although the intensity is reduced. The associated biscattering diagram clearly indicates the loss in intensity, but not in directionality. As any incident wave can be represented as a superposition of plane waves, images will be faithfully reflected. The rescaled medium maintains the principal functionality of the Eaton lens. 

Figure 3 displays our device. We built the transmuted Eaton lens from 10 concentric rings of low--loss printed circuit board (Rodgers$^{{\rm TM}}$ RT6006 with $17.5 \mu\mathrm{m}$ copper coated on one side of $0.127 \mathrm{mm}$ dielectric substrate with $\varepsilon= 6.2 - {\rm i} 0.01$). The rings have a height of $10\mathrm{mm}$ and are spaced with $4\mathrm{mm}$ distance until the radius of the outer ring reaches $40\mathrm{mm}$. The rings carry three rows of split--ring resonators etched out from the copper of the circuit board (lengths $3.14\mathrm{mm}$ and heights $3.33\mathrm{mm}$). The split--ring resonators create the required profile (\ref{rescaled}) of the radial magnetic permeability $\mu_r$ for microwave radiation at $8.9 \mathrm{GHz}$ frequency ($34 \mathrm{mm}$ wavelength). They constitute a metamaterial$^{7-11}$, a material with dielectric properties created by subwavelength structures --- the split--ring resonators in our case. The circuit--board rings were glued on a $0.1\mathrm{mm}$ thick alluminium sheet that was prepatterned with concentric circular prints exactly fitting the rings. Then the rings were filled with dielectric powder (Emerson \& Cuming Microwave Products) to generate the required electric permittivity (\ref{rescaled}) of $n_0^2$. 

Figure 4 shows the curves of the dielectric functions. We designed the split--ring resonators by reverse--engineering where we calculate their dielectric response and modify their parameters until the calculated $\mu_r$  matches the required profile (\ref{rescaled}). However, each split--ring resonator interacts with its neighbors and cannot be considered on its own. For each individual resonator, we calculated the effective dielectric properties using Ansoft HFSS$^{{\rm TM}}$, assuming the resonator to constitute the elementary cell of an infinite lattice$^{22}$ (lattice constants $3.14\mathrm{mm}$, $4.00\mathrm{mm}$ and $3.33\mathrm{mm}$ in accordance with the dimensions of the resonators and the spacing of the circuit--board rings). In this way we embed, in the modeling, the split--ring resonator into a lattice of its own kind. This procedure presumes that the dielectric properties do not vary much and it also ignores the curvature of the circuit--board rings, even at the inner ring of the device where the profile (\ref{rescaled}) contains a topological defect. Nevertheless, our experimental data prove that such a crudely transmuted Eaton lens works remarkably well in practice.

Figure 5 shows the results of our measurements. Our microwave source is a vector network analyzer (VNA) (HP8722D, Agilent Technologies) that we also use for detection. It has two ports connected by two coaxial cables (Megaphase 1GVT4), the feeding and the detecting port. To confine and scan the microwave radiation, we apply the parallel--plate electromagnetic field mapping system described elsewhere$^{23}$. The microwave radiation is confined in horizontal direction by two large alluminium boards at $11 \mathrm{mm}$ spacing. The field is scanned by a sub--wavelength monopole probe antenna, a thin coaxial cable, inserted through a hole at the center of the top plate. The bottom plate, carrying the device, is mounted on a computer--controlled stage that can move in two dimensions. The total scanning area covers $120 \times 240 \mathrm{mm}^2$ with a step resolution of $1\mathrm{mm}$. On the bottom plate pieces of $10 \mathrm{mm}$ thick microwave absorbing foams were placed to define the incidence channel and suppress scattering at the boundaries. An absorbing sheet segregates the incidence channel from the reflected radiation. The sheet was made of a $0.1 \mathrm{mm}$ thick copper tape sandwiched between two $1.5 \mathrm{mm}$ thick layers of magnetically loaded silicone absorber (ECCOSORB).

As Fig.\ 5a shows, the measured field inside the device is perturbed by the graininess of the metamaterial and its ring structure, but the outgoing wave is in very good agreement with the numerical simulation for a smooth dielectric profile displayed in Fig.\ 5b. In simulations, we even reduced the number of rings to 5, and still observed a good performance of the transmuted Eaton lens. There we approximated the device by 5 equidistant uniform layers with constant dielectric properties (\ref{rescaled}). Our crudely approximated prototype preserves the functionality of the ideal Eaton lens, which indicates that the transmutation of singularities$^{17}$ can be remarkably robust and reliable in practice. As the required dielectric properties (\ref{epsmu}) lie within a finite range, such devices can, in principle, work over a broad range of the spectrum and similar devices$^{15,16,24-28}$ could even operate in the visible.

\section*{Acknowledgements}

Y.G.M and C.K.O. are supported by Defense Science and Technology Agency under the Defense Innovative Research Program, Singapore (DSTA-NUS-DIRP/2004/02), T.T. acknowledges the grants MSM0021622409 and MSM0021622419 and U.L. is supported by a Royal Society Wolfson Research Merit Award.

\section*{Author contributions}
Y.G.M. and C.K.O made contributions to the numerical simulations, device design, implementation and the experiment, T.T. and U.L. made contributions to the theory, U.L. suggested this project and wrote the paper. 


\newpage

\begin{figure}[h]
\begin{center}
\includegraphics[width=15.0pc]{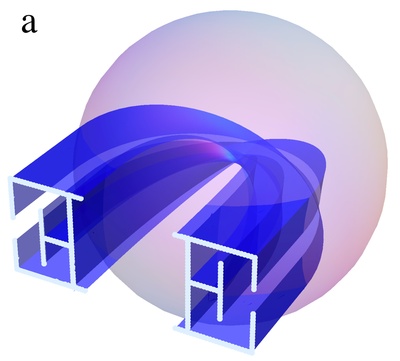}
\includegraphics[width=15.0pc]{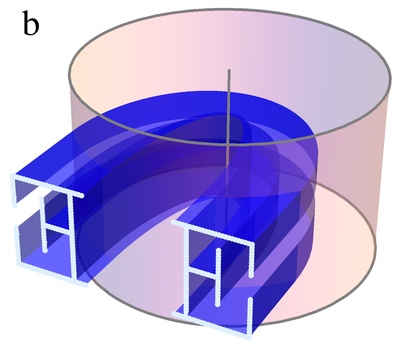}
\caption{\small
{\bf $|$ Eaton lenses.} {\bf a}, Spherical lens. {\bf b}, Cylindrical lens. Artist's impression of the retroreflection of light that carries an image, the letter ``E'' for ``Eaton''. In the outgoing light, the image is inverted, but preserved (in a: flipped and upside down, in b: flipped). The implementation of an Eaton lens would require a singularity in the refractive index profile where the index tends to infinity, unless the singularity is transmuted into a harmless topological defect, as we demonstrate in this paper for the cylindrical lens with metamaterials for microwaves.
}
\end{center}
\end{figure}

\begin{figure}[h]
\begin{center}
\includegraphics[width=35.0pc]{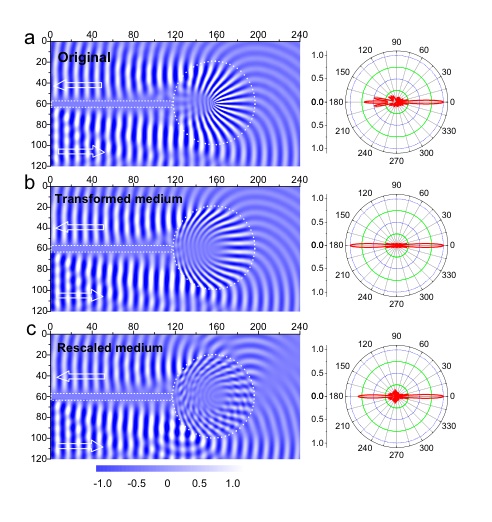}
\caption{\small
{\bf $|$ Simulation of Eaton lenses.} The left pictures show the distribution of the electric field at the original (a) and transformed Eaton lenses (b) and (c) with dielectric functions (\ref{cylinder}) and (\ref{rescaled}), respectively; the right pictures show the corresponding biscattering diagrams. The dashed circles in the left picture mark the boundary of the device at radius $a$ ($40\rm{mm}$), the dotted lines refer to an absorbing sheet that separates the incident and reflected electromagnetic waves with their direction indicated by the dotted arrows. The wavelength is $a/4$ and the scale is in mm. The biscattering diagrams refer to the electric field infinitely far away from the device. They display the ratio of the magnitude of the electric field as a function of angle, normalized by the largest value.}
\end{center}
\end{figure}

\newpage

\begin{figure}[h]
\begin{center}
\includegraphics[width=30.0pc]{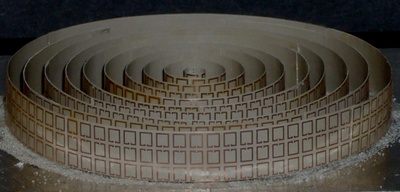}
\caption{\small
{\bf $|$ The device.} The split-ring resonators constitute a metamaterial for microwave radiation with the designed radial magnetic permeability (\ref{rescaled}). The rings were then filled with a white dielectric powder that generates the required electric permittivity of $n_0^2$ (not shown here).
}
\end{center}
\end{figure}

\begin{figure}[h]
\begin{center}
\includegraphics[width=25.0pc]{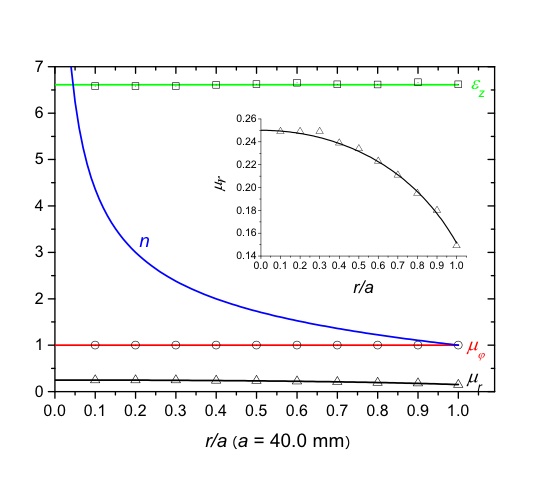}
\caption{\small
{\bf $|$ Effective electromagnetic properties.} The blue curve shows the index profile (\ref{eaton}) of the original Eaton lens as a function of the radius $r$, the other curves display the rescaled dielectric functions (\ref{rescaled}) of the transmuted Eaton lens. The inset shows $\mu_r$ with magnified scale. The symbols are the dielectric functions calculated for the corresponding layers of the metamaterial.
}
\end{center}
\end{figure}

\newpage

\begin{figure}[h]
\begin{center}
\includegraphics[width=30.0pc]{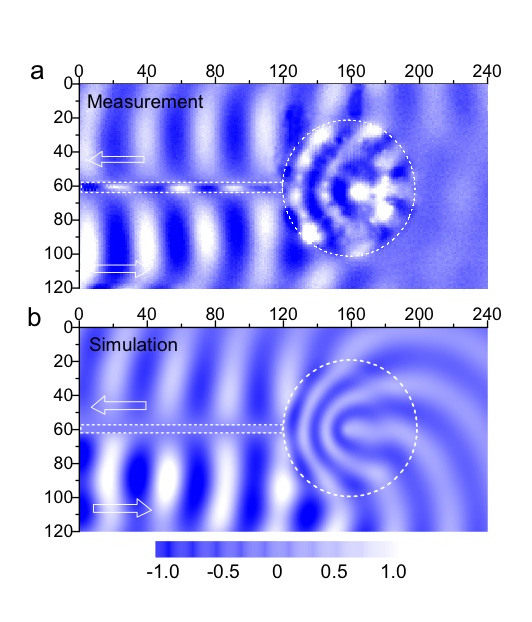}
\caption{\small
{\bf $|$ Measurement results (a) compared with simulation (b)}. Description as for the left pictures of Fig.\ 2, except that here the wavelength is $34 \mathrm{mm}$ ($0.85a$).
}
\end{center}
\end{figure}

\end{document}